\documentclass[manuscript,screen]{acmart}
\AtBeginDocument{%
  }

\begin{document}

  \title[Thick Alignment]{Thick Alignment: A Social Framework for Values and AI Systems}

\author{Alondra Nelson}
\orcid{0000-0003-0383-1806}
\affiliation{%
  \institution{Institute for Advanced Study}
  \city{Princeton}
  \state{New Jersey}
  \country{USA}
}

\author{Sola Kim}
\orcid{0000-0002-2293-9348}
\affiliation{%
  \institution{Institute for Advanced Study}
  \city{Princeton}
  \state{New Jersey}
  \country{USA}
}

\renewcommand{\shortauthors}{Nelson \& Kim}

\begin{abstract}
Computer science research on the ``alignment'' of artificial intelligence (AI) systems treats human values as content to be specified and optimized. Decades of social science scholarship describe values differently, as enacted in ordinary practice, interpreted in social context, and contested across plural orders of evaluation and worth. We argue that the dominant alignment techniques, which we call ``thin alignment,'' convert situated moral or value judgments into one or a few metrics via devices of commensuration such as preference labeling, reward modeling, and consumer ratings. Drawing on philosophical and anthropological accounts of thickness, ``thick alignment'' instead examines how values take shape, inform the development of AI systems, and become subject to contestation over whose interests they serve. We describe two forms of misalignment in seemingly aligned algorithmic systems that highlight thin alignment: ``Halcyon misalignment,'' which presumes a prior, just equilibrium of technology and values from which a system has departed, and ``Mirrored misalignment,'' which names the risks and harms produced by systems that are aligned in a thin sense yet accurately reproduce an unequal social order. A thick account registers such systems as misaligned. We then trace how researchers have taken up and adapted the concept of thick alignment since it was first proposed in 2023, using these adaptations to refine the concept, and to distill three of its pillars: (1) contextual description that both reflects and constitutes values; (2) interrogation of values across the life cycle of a system; and (3) shared, accountable decision-making through which affected publics help determine which values guide the system. We close by considering implications for AI research and policy.
\end{abstract}




\maketitle

\section{Introduction}

On April 25, 2025, OpenAI released an update to GPT-4o, the model behind ChatGPT. Soon after its release, users circulated screenshots on social media of chats with the model in which it lavished praise on banal ideas and agreed with incorrect ones. There were accusations that the updated model had encouraged delusions and suicidal thinking. On April 28, the company pulled back this update in response to public concern and controversy. OpenAI subsequently published an account of what had gone wrong, explaining that it trains its models partly by incorporating user feedback signals, such as thumbs-up and thumbs-down responses to GPT-4o outputs. The update had prioritized this kind of immediate feedback and also did not fully account for how users' interactions with the product might evolve over time. The result, OpenAI argued, was a model that skewed toward ``sycophancy''---overly confirming and supportive outputs~\cite{openaiSycophancyGPT4oWhat2025}.

Anthropic researchers had already documented sycophancy across large language models. A study of five deployed AI assistants---Claude-1.3, Claude-2.0, GPT-3.5-turbo, GPT-4, and Llama-2---found that the human preference labeling used to train these models more positively rated responses that agreed with the user, sometimes over correct responses~\cite{sharmaUnderstandingSycophancyLanguage2024}. Based on this comparative research, Sharma et al. hypothesized that sycophancy may ``be a property of the way these models were trained, rather than an idiosyncratic detail of a particular system.'' The reports of alleged delusion, addiction, and self-harm that followed the release of the GPT-4o update lend support to the finding that models trained on human preferences tend to agree with users~\cite{cuthbertson2025chatgpt,hillLawsuitsBlameChatGPT2025}.

In both the model testing environments and the chatbot interface, thumbs-up or thumbs-down buttons serve as feedback, as signals of users' assessment of AI outputs. But what does a human model tester or consumer mean to convey when they click on one of these buttons?~\cite{movvaWhatsMyHuman2025} The gesture might signal agreement, courtesy, or impatience to move on. It may just be a habit learned and carried forward from the ``like'' buttons on social media or messaging apps, where thumbs-up connotes a litany of meanings from ``message received'' to assent to ``I'm ok with that, even if I don't agree.'' Once the digital gestures are encoded as a discrete feedback signal, these context-dependent meanings are lost and the gesture is translated into quantifiable data~\cite{albrechtDiscerningAudiencesButtons2024,gerlitzEconomySocialButtons2013}. While the thumbs up conventionally denotes the same thing across social media, messaging apps, and chatbot interfaces, what the gesture connotes in each instance varies and requires some effort to interpret beyond a broad sense of agreement. Buttons are not the only feedback signal: how a user responds in the next turn---agreeing, pushing back, rephrasing---also feeds into training~\cite{sharmaUnderstandingSycophancyLanguage2024}, and raises the same problem of interpretation.

The gap between what a gesture denotes and what it connotes is an old problem of interpretation.The philosopher Gilbert Ryle illustrated it with two boys who contracted the eyelids of their right eyes: the movements are identical, butfor one boy, it is an involuntary twitch and, for the other, a conspiratorial signal to a friend. Ryle used the terms ``thin'' and ``thick'' to distinguish between these ways of describing an action~\cite{ryleThinkingThoughtsWhat1971}. Anthropologist Clifford Geertz subsequently deepened this distinction in his account of cultural interpretation~\cite{geertz1973thick}.

Sociologist of science and technology Alondra Nelson brings the problem of interpretation and values to the questions of AI alignment. ``Alignment'' refers to the effort to build AI systems with behavior that accords with human intentions and values~\cite{christianAlignmentProblemMachine2020}. Nelson described dominant approaches to alignment research, of the kind that would later shape models such as GPT-4o, as ``thin alignment'' and contrasted them with the ``thick alignment'' she advocates~\cite{nelsonThickAlignment2023a,nelsonThickAlignment2023}. Thick alignment, as we use the term, attends to how values are interpreted, constituted, and contested in context, rather than treating them as specifications to be met. The point extends beyond how we interpret individual actions. Just as a gesture can mean different things depending on the circumstances, human values cannot be understood apart from the social contexts in which people express and negotiate them. This argument builds on sociological accounts in which values acquire meaning through social processes of interpretation, negotiation, and contestation, becoming embedded in institutions and practices~\cite{lamontComparativeSociologyValuation2012}.

Nelson argues that the approaches to AI alignment often place primacy on technique and in specifying objectives and training models against them. Values, from this perspective, are inputs that already exist externally and await being met or achieved as a part of the AI training pipeline. Technical alignment of AI systems remains a challenge---related to both profit motives and engineering capabilities---as recent public incidents involving AI agents breaching cybersecurity or other organizations have shown~\cite{huggingface2026security}. Yet the values side of alignment is also a hard problem. Societies have long argued over, bargained about, legislated, and fought over values and continue to do so. The alignment approach in AI research therefore requires a much ``thicker'' conceptualization. 


Moving these insights into the discussion of the GPT-4o model, we propose that a system trained to reward thumbs-up records each gesture as general approval, without distinguishing what it denotes in a particular interaction or what it connotes about the users' intentions. By the standard it was trained against, GPT-4o worked as originally intended, optimizing for the feedback it received and adjusting accordingly, without necessarily capturing the intentions or values that users meant to convey. The GPT-4o model was aligned, thinly.

This paper elaborates Nelson's arguments at greater length and engages with the scholarship that has taken them up. We make four contributions. First, we bring theoretical concepts of thickness into dialogue with alignment research, showing how alignment techniques operate as commensuration devices that convert socially situated value judgments into a single or a few metrics. Second, drawing on scholarly literature ,we show examples of thin alignment and identify the related concepts of \emph{halcyon misalignment} and \emph{mirrored misalignment}. Third, drawing on how subsequent research has adapted the concept, we distill three interconnected pillars of thick alignment: contextual description that both reflects and constitutes values; interrogation of values across the life cycle of AI systems; and shared, accountable decision-making through which affected communities help determine which values guide them. Finally, we consider the implications of thick alignment for AI research and policy.

\section{The Alignment Problem}

In a 1960 essay in \emph{Science} on the moral and technical consequences of automation, the cybernetician Norbert Wiener warned that if we use a mechanical agent ``with whose operation we cannot efficiently interfere,'' then we ``had better be quite sure that the purpose put into the machine is the purpose which we really desire''~\cite{wienerMoralTechnicalConsequences1960}. Sixty years later, Brian Christian's book \emph{The Alignment Problem}~\cite{christianAlignmentProblemMachine2020} revisited this predicament in the context of contemporary AI development, outlining the gap between the objectives built into human-made technological systems and the fulfillment of human values that AI researchers still confront.

Christian identifies an example of this gap in Joy Buolamwini and Timnit Gebru's research on facial analysis technology. As they detailed in their paper, ``Gender Shades,'' two widely used facial recognition technology reference data benchmarks were overwhelmingly composed of lighter-skinned and male subjects. Buolamwini and Gebru evaluated three commercial gender classification systems on a new benchmark they devised that included more equitable representation of both skin tone and gender. The accuracy of these systems ranged from 88 to 94 percent overall, but when the researchers disaggregated the results, darker-skinned women were the most misclassified group, with error rates as high as 34.7 percent, while the maximum error rate for lighter-skinned men was 0.8 percent~\cite{buolamwiniGenderShadesIntersectional2018}. Prior to this work, technology company developers had assumed, with some justification, that the facial classification systems were aligned with the objective of classifying faces to some degree of accuracy against labeled data. However, in practice, Buolamwini and Gebru's research shows that technical systems execute an objective that encodes a choice --- in this case, about whose faces count as the default. That choice is both value laden and socially consequential.

Large language models present a different alignment problem. With facial analysis, the alignment objective was presumably simple and was to classify faces to some degree of accuracy against labeled data, with values treated as given or externalities. For most prompts to a language model, by contrast, the correct output is not specified. With generative AI, the question that the ``Gender Shades'' researchers exposed---which faces define good system performance---became a question about what counts as a good response. Developers at technology companies answered in part by asking people---testers and users---which responses they preferred, with values trending to the mode or to the judgment most raters shared.

Large language models are aligned using reinforcement learning from human feedback or RLHF~\cite{christianoDeepReinforcementLearning2017,ouyangTrainingLanguageModels2022}. Human raters are shown pairs of model outputs and asked which they prefer. These comparisons are then used to train a reward model, a separate model that learns to predict raters' preferences and assigns a number to each output. The LLM is then trained to produce outputs that are scored highly by the reward model. Constitutional AI, another alignment approach developed by Anthropic, complements some of the human ratings with a list of principles that the model is instructed to align with to revise its outputs~\cite{anthropicClaudesConstitution2026,baiConstitutionalAIHarmlessness2022}. With this approach, alignment to human values also means alignment to a company's idea of what is good for people. User ratings, such as thumbs-up and thumbs-down buttons in products such as ChatGPT or Claude are another source of feedback. Each of these techniques converts judgment, whether of raters and users or of the company itself---and with it human values---into a signal against which a model is trained and aligned.

Such AI alignment techniques are commensuration devices. Sociologists Wendy Espeland and Mitchell Stevens define ``commensuration'' as the transformation of different qualities into a common metric, such as prices and rankings. They show that commensuration is a social process with consequences. It simplifies and reduces social context, discarding information and redirecting attention toward the outcomes the metric captures and reifies. Commensuration also redistributes authority toward those who define a metric and away from those who contribute data to it or are subject to it~\cite{espelandCommensurationSocialProcess1998}. A reward model, for example, takes judgments about harm, helpfulness, honesty, offense, or other reactions---judgments that, in ordinary life, are made by different people for different reasons in varied contexts---and converts them into a metric. The conversion is useful for the training of AI models. It also performs two moves discussed by Espeland and Stevens. First, the process discards the reasons a rater had for preferring one response over another. Second, it directs the developer's attention toward ratings as actionable information and away from the contextual situations to which the ratings responded. Through both moves, authority is conferred on and transferred to whoever creates the rating framework.

Commensuration helps explain what the conversion reshapes value judgments, but not their social origin. Michèle Lamont's sociology of valuation and evaluation addresses this gap by showing how value is generated through acts of categorization and how the criteria people apply make sense only within particular evaluative cultures~\cite{lamont2009professors,lamontComparativeSociologyValuation2012}. Read through Lamont's framing, preference labeling to train AI models lifts judgments from the settings in which raters ordinarily make them.For example, when OpenAI hired about forty contractors to rank model outputs in training InstructGPT, a precursor to ChatGPT, it selected them through a screening test and guided them with written labeling instructions. The developers acknowledged that the resulting model was aligned to the preferences of these labelers as shaped by the company's instructions~\cite{ouyangTrainingLanguageModels2022}. The reasons one rater found one output better than another were not recorded in the training signal, but arguably became encoded in the model. Commensuration, in this light, is also a process of values generation.

Evaluative cultures differ across people and places, and two recent studies in machine learning research document what that variation means for alignment. Shibani Santurkar and colleagues set out to learn whose opinions language models express~\cite{santurkarWhoseOpinionsLanguage2023}. They posed questions from the Pew Research Center's long-running surveys of American public opinion to a set of language models and compared the answers with those of sixty demographic groups. The distance between the models and many of these groups was substantial, and it changed with alignment. AI models refined with human feedback came to resemble some groups far more than others, drifting toward the views of respondents who were liberal, higher-income, college-educated, and nonreligious. Whereas Santurkar and colleagues measured the outputs of presumably aligned models, the PRISM project, led by Hannah Rose Kirk, examined the feedback that goes into them~\cite{kirkPRISMAlignmentDataset2024}. The research team recruited fifteen hundred participants from seventy-five countries, documented their demographic backgrounds and stated expectations of AI systems, and traced every rating they gave to twenty-one language models back to the person who gave it. The backgrounds of the participants shaped the engagements they chose to have, especially on value-laden subjects such as abortion and Middle East politics. The study also found disagreement about which responses were best, and which models ranked highest depended on whose ratings were counted. The values at stake in alignment, and treating them as settled inputs to a technical process favors some people's judgments over others.

Together, these studies show that the values at stake in alignment as settled inputs to a technical process favors some people's judgments over others. Recall Wiener cautioned that the purpose put into the machine had better be the purpose ``which we really desire.'' Thin alignment converts judgments into one or a few metrics, and in doing so settles, in advance and largely out of view, whose judgments will count, and who Weiner's ``we'' includes. When someone rates a model's outputs, whether a contracted tester comparing responses or a user clicking thumbs-down, the reasons for the choice disappear and the choice is carried forward as a metric with authority.

\section{The Implications of Thickness for Alignment}

Thick alignment draws on an intellectual genealogy of ``thickness'' in the humanities and social sciences that distinguishes describing what people do from understanding what their actions mean. A thin account abstracts from context. A thick account keeps hold of the context that gives an act or a concept its meaning. The distinction began as a claim about how to describe actions, became a claim about the concepts people use to judge them, and then became a claim about how people disagree. Table~\ref{tab:thinthick} summarizes the contrast between thin and thick conceptions of alignment that follows from this account.

\begin{table}[!htbp]
  \caption{Thin and thick alignment compared.}
  \label{tab:thinthick}
      \begin{tabular}{p{0.16\linewidth} p{0.36\linewidth} p{0.40\linewidth}}
        \toprule
        \textbf{Dimension} & \textbf{Thin alignment} & \textbf{Thick alignment} \\
        \midrule
        Unit of analysis & Model output and its rating & Situated interaction and its context of use \\
        Source of values & Rater preferences, user ratings, and written principles & Practices and institutions in which values are made and contested \\
        Site of difficulty & Specification and optimization & Moral disagreement and interpretive authority \\
        Evaluation & Benchmark scores and preference win rates & Validated measurement joined to interpretive and comparative inquiry \\
        Temporal scope & Training and pre-deployment testing & The full life cycle, including use and revision \\
        Who decides & Developers and contracted raters & Developers with affected publics, through shared, accountable decision-making \\
        \bottomrule
      \end{tabular}      
\end{table}

We follow that lineage to draw out three claims about human values, each of which has consequences for how AI alignment is practiced and evaluated: (1) Values are interpreted; that is the same act can carry different values depending on who performs it and under what shared codes; (2) Values are constituted by which we mean they take shape in the concepts and relationships through which people judge one another and, therefore, they cannot be fully specified in advance of those relationships; (3) Values are contested in that people disagree about them and the work of resolving those disagreements traditionally sits with social and political processes and institutions.

\subsection{Values Are Interpreted}

The first claim comes directly from the genealogy of thick description. Ryle pressed his example of the two boys, one of whom twitches while the other winks. The same contraction of the eyelid might also be a parody of someone else's clumsy wink or the parodist rehearsing that parody in front of a mirror~\cite{ryleThinkingThoughtsWhat1971}. The movements are indistinguishable. For Ryle, an adequate description of these movements must capture what the actors are doing, layer upon layer of intention in context.

Geertz made Ryle's distinction between thin and thick description the basis for innovation in social science methods and in doing so he moved its weight from the individual's intention to sociality. A wink, Geertz observed, communicates ``according to a socially established code''~\cite{geertz1973thick}. In ``Thick Description: Toward an Interpretive Theory of Culture,'' the opening essay of \emph{The Interpretation of Cultures}~\cite{geertz1973thick}, he argued that the work of ethnography consists in sorting out the structures of signification through which behavior becomes meaningful. His essay on the Balinese cockfight, ``Deep Play,'' illustrates the point. A thin account would describe two roosters fighting and men placing bets. Geertz's account read thickly shows that the betting follows lines of kinship and status, so that the fight dramatizes the social hierarchy of the men who wager on it and offers a window onto the social order more broadly. A rater's thumbs-up, like a wink, communicates according to a code that the rating itself does not record.

Thick description is not without its critics. In \emph{Thin Description}~\cite{jackson2013thin}, John L. Jackson Jr. confronted the god's-eye perspective implicit in Geertz's approach. He challenged the assumption that an ethnographer, by dint of time in the field, can achieve some comprehensive and privileged understanding of a community. Jackson's own study of the African Hebrew Israelites of Jerusalem takes a different approach, emphasizing that every observer, himself included, sees a community from a particular and partial view point. Heather Love made a complementary argument in ``Close Reading and Thin Description''~\cite{loveCloseReadingThin2013}. Drawing on the sociology of everyday interaction, especially the work of Erving Goffman, she defended careful, surface-level observations of behavior as important. Thus thick description is not necessarily a total or authoritative one.

These arguments have implications for AI alignment. A thin account of an interaction with an AI system records that a model produced an output and that a rater or user did or did not prefer it. A thick account asks what kind of exchange took place, who was party to it, which social codes governed the input and also how the output was read. Jackson's critique reminds us that those who build such an account, developers and researchers among them, see from some vantage. The people who use AI systems, and those affected by them, can say what an exchange meant in terms an outside observer would miss. Love's defense of thin description suggests that the surface of an interaction, such as a chatbot user's thumbs-up, may be read carefully. Thick alignment accounts for these feedback interactions alongside accounts of how they were constituted and of the evaluative settings in which they were made.

\subsection{Values Are Constituted}

Bernard Williams extended the idea of thickness to the concepts people use to judge actions. In \emph{Ethics and the Limits of Philosophy}~\cite{williamsEthicsLimitsPhilosophy1985}, he distinguished thin moral concepts, such as good and right, from thick ones, such as cruelty, courage, betrayal, and gratitude. Thick concepts, Williams argued, do more than describe the world; they also shape how people respond to it. To call an act cruel, for example, is to describe what happened, perhaps that someone caused another person needless suffering, but it is also to condemn the act. The two judgments cannot easily be separated. A person who has learned the concept of cruelty has learned both what to recognize in an action and what about it should be deplored.

Williams's distinction helps explain what happens in AI alignment. Rating guidelines ask human raters to judge model outputs using categories such as harmful, helpful, honest, and respectful. These are thick concepts. Applying them requires more than identifying a feature of a response; it means making judgments about the situation and about what matters in it. Raters bring their own moral understandings to these decisions. But preference labeling reduces each judgment to a single verdict. The reasoning behind that judgment, including what the rater understood by harmful or helpful, does not carry over into the resulting data. Written principles for AI systems face a similar problem. A rule against harmful outputs says little until someone applies it to a particular statement, directed at a particular person or made within a particular relationship. What counts as harmful depends on that encounter, and the meaning of the value is worked out each time someone applies it, whether in testing or in use.

Values also take shape in interrelationships and are enacted in ordinary practices. For example, support means one thing between friends, another between a counselor and a client, and perhaps another between a person and a chatbot. A confidant who agrees with everything we say may be offering kindness; an LLM output that does the same may cause harm. Many of the people who turned to GPT-4o in the spring of 2025 treated it as a confidant~\cite{hill2025confided}. A model trained to curry favor with users could register that a user was pleased, but not that its sycophantic responses put them at risk. In \emph{Animals, Robots, Gods}~\cite{keaneAnimalsRobotsGods2025}, the anthropologist Webb Keane places chatbots within the long human history of moral relations with nonhumans, from oracles and spirits, to gods and animals and he observes that people meet new social objects with interpretive frameworks they already hold. To understand what values someone brings to engagement with technology, Keane argues, one has to know the world that person lives in, from the household and livelihood that organize their days to the tales about machines that circulate around them. Values, on this account, are constituted in concepts and in relationships.

\subsection{Values Are Contested}

If values take shape in particular concepts and relationships, people formed in different ones will understand them differently. In \emph{Thick and Thin}~\cite{walzerThickThinMoral1994} political theorist Michael Walzer brought the concept of thickness to the question of moral disagreement. Viewers around the world, he observes, understood at once signs reading ``Truth'' and ``Justice'' carried by demonstrators in Prague in 1989, yet the meanings the marchers attributed to those words were embedded in their particular history of occupation and dissent. For Walzer the widely shared minimum constitutes a thin morality and the culturally embedded maximum a thick one. The values we understand as morality, he argued, are ``thick from the beginning''~\cite[p.~4]{walzerThickThinMoral1994}, and the thin is abstracted from many thick moralities and comes after them. Thin alignment reverses Walzer's order. It begins with minimal terms such as ``helpful'' and ``harmless'' and treats them as a foundation from which particular judgments can be derived. Walzer's account also helps to show why thick alignment can sustain contestation---there can be many meanings---yet also does not collapse into relativism, or their being no meaning. People can often agree at a certain level of generality, even as they appeal to plural orders of worth, distinct standards for judging what is valuable or fair~\cite{boltanskiJustificationEconomiesWorth2006}.

For AI alignment, Walzer's account suggests where agreement can rest and where it cannot. The fuller meanings of those shared values remain rooted in the particular communities that hold them and are worked out through their own social and political institutions. An AI model used across different communities encounters these different understandings at once. Deciding which values it should reflect, and whose judgments should guide it is a matter of authority and debate as well as design.

If values are interpreted, constituted, and contested in context, then AI alignment is a social practice as well as a technical one. Thick alignment, in this account, attends to the contexts in which values are interpreted, the concepts and relationships through which they are constituted, and the institutions through which disagreements about them are worked out.

\section{The Pitfalls of Thin Alignment}

Thin alignment converts plural, situated judgments into one or a few metrics. When systems aligned in this way cause harm or are used or act in unintended ways, they are commonly described as misaligned. In the technical literature, misalignment names a gap between a system's behavior and the objective its developers intended. The word misalignment implies a departure from a correct state. To call a system misaligned is to suggest that it once worked as it should, or would have, until something pulled it off course. But this diagnosis is rarely applied to systems that meet their technical objectives, such as the facial analysis systems discussed earlier, even when those systems produce bias or harm. Thick alignment changes what counts as misalignment. If values are interpreted, constituted, and contested in context, then a system can meet its technical objective and still be misaligned with the values of the publics it affects. Moreover, a diagnosis of misalignment can itself rest on mistaken perceptions about values and social arrangements against which systems are evaluated.

We name two concepts that clarify what thin alignment misses. The first, ``\emph{halcyon}'' misalignment, names a diagnosis that presumes a prior state of just alignment between technology and society, from which a system has departed. The second, ``\emph{mirrored}'' misalignment, names harm produced when a system meets its technical objectives while reproducing and reinforcing an unjust social order. Both show how a system can be deemed successful or safe according to its own evaluation standards---that is thinly aligned---while remaining misaligned with the values and interests of the publics it affects.

Halcyon and mirrored misalignment are connected because they describe the same baseline from opposite sides: halcyon misalignment assumes that harmful outputs result from a system deviating from an otherwise neutral baseline. Mirrored misalignment shows that the baseline itself may be responsible for the harm. These observations also change how we think about improving these systems. If the underlying data reflects an unjust social order, repair cannot simply mean bringing the system back to some supposedly fair, thinly aligned, state. Instead, correcting misalignment requires questioning and changing the social conditions that produce and are produced by AI systems.

\subsection{Halcyon Misalignment}

The word misalignment implies a departure from a correct state. To call a system misaligned is to suggest that it once worked as it should, or would have, until something pulled it off course. Nelson described as the assumption of a prior ``\emph{halcyon}'' state, an equilibrium in which an AI system and the social world it serves were deemed to be in good order or aligned~\cite{nelsonThickAlignment2023a}. The adjective recalls halcyon days, remembered as calm and untroubled, and often remembered more fondly than they were lived. We call a diagnosis of misalignment that presumes a prior just alignment of technology and values, from which a system has departed, halcyon misalignment.

The trouble with the halcyon assumption is that the baseline is not always just or fair. A diagnosis of halcyon misalignment treats the model as the source of the harm and discounts the historical values and assumptions that shaped its training. Training data are an archive of past social choices: who was hired, who was lent money, who was arrested, whose face was photographed and labeled, whose intellectual property was deemed free to use, whose language became the lingua franca. A model trained on that archive learns the values enacted in the past, which may differ sharply from the values a society or community professes in the present or aspires to in the future.

The use of hiring algorithms offers a clear case. Interpreted as misaligned, these systems appear to have strayed from a fair hiring process; described thickly, they reproduced patterns already present in the hiring process. In 2018, Amazon abandoned an experimental automated resume-screening tool that had learned, from a decade of mostly male applicants' resumes, to downgrade resumes containing the word ``women's.''~\cite{dastin2018amazon}. The same pattern appears in the language models now marketed for screening job applicants. Kyra Wilson and Aylin Caliskan asked three open-source language models to rank more than 550 real resumes against more than 500 job listings, changing only the names on the resumes~\cite{wilsonGenderRaceIntersectional2024}. The models favored white-associated names in eighty-five percent of comparisons and Black-associated names in nine percent of them. In no instance was a supposed Black man's name preferred over that of a supposed white man.

For decades, field experiments and employment audits that sent employers matched resumes differing only in the applicant's name have measured racial discrimination in hiring in the U.S. A meta-analysis of those studies found that white applicants received, on average, thirty-six percent more callbacks than equally qualified Black applicants, with no appreciable change since 1989~\cite{quillianMetaanalysisFieldExperiments2017}. This historical record complicates the idea that these systems departed from an otherwise fair process.

\subsection{Mirrored Misalignment}

Mirrored misalignment describes what that explanation fails to see, which is the harm that comes from a system working as designed. We use this term to characterize harm produced by a system that is aligned in a thin sense, meeting the objective its developers set, while reproducing an unjust social order. Such a system is aligned with the values enacted in the data and practices from which it is built, but misaligned with the values a society or social group professes or aspires to; it mirrors the unjust world as it is, translating its inequalities into technical outputs that reproduce and reinforce them. The point is not that the system deviated from a neutral baseline but that no neutral baseline existed to align to.

A well-documented case of mirrored misalignment is COMPAS, a risk-assessment algorithmic system that courts in several states have used to estimate the likelihood that a defendant will be rearrested; judges have consulted its scores in decisions about bail and sentencing. In 2016, journalists at \emph{ProPublica} analyzed its scores for defendants in Broward County, Florida~\cite{angwinMachineBias2016}. Black defendants who were not rearrested over the following two years were nearly twice as likely as white defendants to have been labeled higher risk. White defendants who were rearrested were more often labeled low risk. Northpointe, the company that developed COMPAS, responded that its scores were equally predictive across groups, meaning that a given score meant roughly the same likelihood of rearrest for a Black defendant as for a white one.

Both of the company's claims were true as subsequent researchers revealed. Prompted by the COMPAS debate, two research teams showed in 2016 and 2017 that no risk score can be fair in every sense at once when two groups are rearrested at different rates~\cite{chouldechovaFairPredictionDisparate2017,kleinberg2017inherent}. If the label is equally accurate for everyone, more people in the group with the higher rate will be wrongly flagged. If the mistakes are spread evenly, the label stops meaning the same thing from one group to the next. The COMPAS tool met the criteria the company had chosen and this social choice then determined who would bear the cost of its errors.

In this sense COMPAS rendered a world organized by inequality with considerable fidelity. The deeper problem lies in what the tool was asked to predict. COMPAS predicts rearrest, and arrest records reflect where police patrol and whom they stop as much as who breaks the law. A model trained on those records learns the practices of policing and returns them as ``predictions'' about individuals~\cite{greenFalsePromiseRisk2020}. The sociologist Ruha Benjamin calls technologies of this kind the ``New Jim Code'' referring to systems that reproduce existing inequities while being promoted as more objective than the discriminatory practices they replace~\cite{benjaminRaceTechnologyAbolitionist2019}.

The same pattern extends to generative AI systems. A language model refined on human preferences mirrors the preferences of the people whose ratings it learns from~\cite{santurkarWhoseOpinionsLanguage2023,fleisigPerspectivistParadigmShift2024,kirkPRISMAlignmentDataset2024}. The sycophancy described in the introduction is one form of this reflection in which raters and users rewarded agreement and the model was trained to agree. In each case, the model follows its training, reflecting back a narrower world than the one in which the product is deployed and used.

\section{Thick Alignment in Practice}

The two forms of misalignment illustrate why technical performance alone is insufficient for understanding alignment. But what does a thicker understanding of alignment require in practice? Since the concept was first proposed in 2023, it has been taken up in computer science papers and preprints, in anthropological essays, in policy reports on AI evaluation and safety, and in medical research. Reading across this work, we can see where the meaning of the concept has shifted and how it has been put into practice. These engagements have not produced a single, settled theory of thick alignment; they are overlapping themes that we distill into prevailing emphases. Notably, each emphasis responds to one of the claims about values set out above: that values are interpreted, constituted, and contested. Thick alignment comprises (1) contextual description: sociotechnical approaches to the analysis of AI models and systems that attend to use cases, incentive structures, power, geopolitics, meanings, and communities, and in so doing both reflect and constitute human values; (2) interrogation of values across the life cycle of an AI model or system, including its potential benefits and harms; and (3) shared, accountable decision-making through which affected communities help determine which values guide the development and deployment of AI models and systems. These emphases are related rather than mutually exclusive. Contextual description examines how values acquire meaning; life-cycle interrogation traces how they are introduced and transformed; and shared decision-making addresses who has the authority to determine which values prevail.

As researchers have engaged with thick alignment, the meaning of thickness has shifted. In much of this work, thick alignment serves as a name for the social dimensions of alignment that technical methods tend to set aside. Some researchers pair Nelson's lectures~\cite{nelsonThickAlignment2023a,nelsonThickAlignment2023} with a related contrast drawn by the sociologist Jacob Foster. In ``From Thin to Thick''~\cite{fosterThinThickPolitics2023}, Foster distinguishes a thin approach to building AI, constructed from the top down and meant to apply everywhere, from a thick approach, in which AI is co-constructed locally and remains open to dispute. Jha, Svarc, and Bagiński~\cite{jhaPositionAIAlignment2026} observe that concerns about AI systems amplifying unjust power structures have ``found explicit conceptualization'' under the heading of thick alignment, marking out territory that alignment research has neglected.

In much of the technical and safety literature, thickness has come to describe a quality expected of the AI system itself, rather than the way people understand and describe the system in its social context. It becomes something an AI system should possess: a contextual understanding of human values~\cite{alikhaniHypeHarmWhy2025}, norms a model should internalize~\cite{stanczakSocietalFrameworksCan2026}, or a grasp of ``the context-sensitive and thick dimensions of human values''~\cite{jhaHumanAmplificationIntelligent2026}. Jha, Svarc, and Bagiński~\cite{jhaPositionAIAlignment2026} paraphrase Nelson as asking that deployed systems ``both reflect and constitute human values.'' Thickness becomes one more quality to be specified and measured, and in the process it is thinned. The thinning happens through operationalization: once thickness is written into a rubric or a benchmark, it becomes a score. Other researchers attach thickness to the moral quality of what a system does. Hammond et al.~\cite{hammondMultiAgentRisksAdvanced2025} use ``thick'' for alignment that includes the idea that a system does what is good or beneficial, and Henselmans~\cite{henselmansThinAlignmentCant2025} calls for systems attuned to ``the thick weave of ethical life.'' These readings reflect a familiar tendency in technical fields to treat desirable qualities as things that can be built into systems and measured. But the readings also bring thick alignment back into the very framework it was meant to challenge, one in which human values become specifications to be measured, and so defeat the purpose of the concept.

The work that engages the concept most fully keeps thickness as an account that people give of a system and its setting. Schmer-Galunder~\cite{schmer-galunderAnthropologyAI2025} returns to Geertz's cockfight to show that alignment can satisfy stated preferences while ignoring ``the cultural embeddedness that makes those preferences meaningful in the first place.'' Schwartz and Waters~\cite{schwartzWildThickWicked2026} build an approach to evaluation on the situated accounts of people who test AI systems. Brandt et al.~\cite{brandtEngagingAdvisoryBoard2025} bring the concept into deliberations with an advisory board for a clinical decision system. In each case, thickness describes what people do with and around a system: how they interpret it, how they follow it through its uses, and how they decide what it should value. Rather than treating thickness as a quality that can be built into a model, these approaches attend to the relationships, practices, and judgments through which the meaning and consequences of a system take shape.

\subsection{Contextual Description}

The first emphasis answers the claim that values are interpreted. A rating or a benchmark score means little until someone describes the setting in which it was produced. Values are made in social relationships~\cite{keaneAnimalsRobotsGods2025} and through incentives and power, and they change as those relationships and incentive structures change. An AI system therefore cannot be held up against them like a ruler or a litmus test. The ways people come to know the world and the ways they order it are produced together. An AI system reflects the values of the data it is built upon, the developers who built and trained it, the feedback it receives, and the context into which it is deployed.

Kirk et al.~\cite{kirkWhyHumanAI2025} make a version of this case in arguing that human--AI relationships require what they call ``socioaffective'' alignment. They call for moving from ``narrow, assumption-ridden or `thin' specifications of alignment'' toward a thick description that ``examines the deeper contexts and layers of meaning in which AI systems operate.'' A description of this kind shows how an AI system is shaped by the social structures into which it is deployed, and how the system in turn reshapes those structures through feedback loops. Schmer-Galunder~\cite{schmer-galunderCultureCodeAnthropological2026} reads thick alignment as naming approaches that use contextual analysis to understand how human values emerge and what they signify. This mutual shaping is co-production in AI development~\cite{fosterThinThickPolitics2023,jasanoffStatesKnowledgeCoproduction2004}. The fullest applications of contextual description concern evaluation.

Schwartz and Waters~\cite{schwartzWildThickWicked2026} propose that the reports of people who test AI systems in realistic settings serve as primary evidence about how those systems behave. The testers are not scored or optimized against a target. Their accounts describe system behavior in context: who did what, where, and with which system. That situated detail, the authors argue, reduces the ambiguity of terms like ``harmful'' or ``unsafe'' and clarifies what they mean in practice for specific deployments. We might read their proposal as putting Bernard Williams's insight to work. A thick concept such as harm acquires its meaning in application, and contextual description shows how that application unfolds. Bogen~\cite{bogenAssessingAISurveying2026} makes a related case for AI assessment. Focused, holistic inquiry, she argues, can supply the contextual analysis that thick alignment requires. It can also strengthen more precise measures by revealing the context and confounding factors they would otherwise miss.

Contextual description also bears on whether values can be deliberately encoded in technology at all. Drawing on Nelson~\cite{nelsonThickAlignment2023a}, Serrano, Brumbaugh, and Smith~\cite{serranoLanguageModelsGuide2023} conclude that the final adjustments made to a language model's behavior are better understood as customization than as the encoding of human value. Artifacts carry political choices, whether or not their designers intend them to~\cite{winnerArtifactsHavePolitics1980}. Contextual description makes those choices visible, so that the people who live with a system can understand and challenge them.

Description of this kind also calls for a sense of proportion. AlphaFold's protein-structure predictions primarily inform scientific inquiry; the same technology enters the development of a diagnostic or a drug the effects of which reach patients more directly and can shape decisions about their care. The need for thick alignment grows as AI systems become involved in decisions with direct consequences for people's health, opportunities, and livelihoods.

\subsection{Values Across the Life Cycle}

The second emphasis answers the claim that values are constituted. If a value acquires its meaning each time someone applies it, in model testing or in AI use, then alignment cannot be settled once and for all when a model is trained. Yet much alignment work focuses on this point in the process, asking which values a model should reflect before it is deployed. But values enter the process much earlier, when training data are selected, and continue to shape decisions about where a system is used and whom it serves. They remain at stake throughout its use. Nelson described this pattern as ``values all the way down''~\cite{nelsonThickAlignment2023a,nelsonThickAlignment2023}, a phrase Smart et al.~\cite{smartTheoryValueAI2026} place at the head of their discussion of alternative conceptions of value in alignment.

At each stage of development, values take on different forms rather than passing unchanged through the AI stack. They become categories in datasets, judgments in annotation guidelines, objectives in model training, measures in evaluation, and decisions about deployment. In moving from one stage to the next, some meanings are retained while others are lost or obscured. These translations also affect whose judgments carry weight. A value that seems clear at one stage may become uncertain or contested when it is applied at another. Thick alignment therefore requires attention to the values that guide a system at a particular moment. It also turns our attention to how those values are interpreted and transformed as an AI system moves from development into use.

Research on evaluation shows how hard this emphasis is to put into practice. In the design document for a National Institute of Standards and Technology program on the risks and impacts of AI, Schwartz et al.~\cite{schwartzAssessingRisksImpacts2024} treat contextual information as a form of thick description. They find it difficult to integrate across the life cycle of a system, because machine learning relies on lightweight, process-based techniques built for production and scalability. Benchmarks, the field's dominant evaluation tool, cannot account for how people use, interpret, and act on AI outputs once a system is deployed~\cite{schwartzRealityCheckNew2025}. Context gathered at one stage of a system's life is often lost at the next. It goes missing most often after deployment, which is where the consequences unfold. The mirrored harms of the COMPAS algorithm detailed above came into view only when journalists followed its scores into the courts where they were being used in judicial decision-making.

Researchers have developed documentation practices that help keep this context from being lost, at least in part. Datasheets for datasets document the provenance and intended uses of training data~\cite{gebruDatasheetsDatasets2021}. Model cards document a model's intended uses and its performance across groups~\cite{mitchellModelCardsModel2019}. Both treat documentation as an ethical practice and a sociotechnical one. These instruments could be extended to document the value choices made at each stage. A model card, for example, could be leveraged to note which taxonomy of harm a developer adopted, who wrote the rating guidelines, who the model raters were, where they worked, and how disagreements among their ratings were resolved. Such documentation would not resolve disagreements about values, but it would make the choices and judgments behind a system easier to see, question, and challenge.

\subsection{Shared, Accountable Decision-Making}

The third emphasis answers the claim that values are contested. People disagree about values, yet every AI system embodies some of them. Whenever a system is built, a decision about whose values prevail has been made, whether or not anyone acknowledges it. Nelson framed this emphasis around who gets to decide, and who gets to take part in the deliberation through which the values encoded in technologies are chosen~\cite{nelsonThickAlignment2023a}. She examined one attempt to widen that circle. In 2023, Anthropic and the Collective Intelligence Project asked a representative sample of about a thousand Americans to propose and vote on principles for a model's constitution. Nelson commended the project's transparency but also cautioned against what she called a turn to ``corporate democracy projects,'' in which a survey of a thousand people, however representative, is intended to stand in for democratic institutions that societies have built over generations~\cite{nelsonThickAlignment2023}.

Researchers who cite thick alignment have attended most closely to the accountable part of this emphasis and to who holds the power to set the terms of deliberation in the first place. The customization that Serrano, Brumbaugh, and Smith~\cite{serranoLanguageModelsGuide2023} describe is performed by a small number of companies. Jha, Svarc, and Bagiński~\cite{jhaPositionAIAlignment2026} place concerns about AI systems ``amplifying unjust power structures'' under the heading of thick alignment. Kirk et al.~\cite{kirkWhyHumanAI2025} count institutions, governance mechanisms, and market power among the contexts a thick account must examine.

Accountability also depends on whom evaluation serves. Schwartz and Waters~\cite{schwartzWildThickWicked2026} observe that the measurement community around AI is oriented toward helping developers build better models, ``not to inform the organizations, practitioner communities, and stakeholders who need to understand what AI means in their own settings.'' Bogen~\cite{bogenAssessingAISurveying2026} treats focused, contextual assessment as ``a crucial dimension of AI assessment and accountability,'' one that can bring to light impacts that existing approaches have not ``sufficiently recognized or prioritized.'' Participation alone, however, does not guarantee shared decision-making. At issue is not only whether affected communities have meaningful influence over the terms of deliberation and the decisions that follow. It also matters whether they can hold decision-makers accountable when the values they have expressed are disregarded.

Brandt et al.~\cite{brandtEngagingAdvisoryBoard2025} offer the clearest case of decision-making that is shared as well as accountable. They invoke thick alignment in their work with an advisory board convened to deliberate about bias and fairness in a clinical decision support system. They describe it as ``the recognition that there are multiple values in tension when it comes to the design of algorithmic systems and the importance of understanding these concerns in context and their relationship to structural inequities.'' Their own interests as researchers, they acknowledge, included gathering more data and prioritizing model performance, and these interests could clash with those of patients. From the board's deliberations, they conclude that a system favoring clinicians' priorities over patient-defined outcomes would likely be unjustifiable, and that unequal power in system design calls for extra care in amplifying the voices of those who are structurally disadvantaged. They are explicit that the stakes include accountability as well as system design. Making clear ``how and why certain parties' values are built into AI/ML systems,'' they write, ``is critical in ensuring that resulting systems are trusted by affected parties.''

\section{Thick Alignment for Policy}

The three emphases of thick alignment have implications beyond the research and evaluation of AI systems. They also concern the institutions that determine how systems are developed, where they are deployed, whose interests they serve, and who can challenge the decisions made about them. Contextual description requires institutions to understand systems in the settings where they operate, rather than assess them against technical measures alone. Interrogation across the life cycle requires arrangements that preserve this understanding as systems move from development into deployment and use. Shared, accountable decision-making requires institutions through which affected publics can influence the values that guide those systems and hold decision-makers answerable for their choices. These are also questions of governance. Policy can establish the conditions under which such work takes place, but it can also reproduce the limitations of thin alignment when it translates complex social judgments into standardized requirements that are treated as sufficient in themselves.

Some existing governance frameworks provide starting points. The National Institute of Standards and Technology's AI Risk Management Framework describes AI systems as ``inherently socio-technical in nature'' and organizes risk management around four functions: govern, map, measure, and manage~\cite{nationalinstituteofstandardsandtechnologyArtificialIntelligenceRisk2023}. It calls for attention to the context of deployment, the people and communities affected, and risks that emerge as systems are used. The European Union's AI Act takes a more specific regulatory approach. Article 27 requires certain deployers of high-risk AI systems, including specified public bodies and providers of public services, to assess potential impacts on fundamental rights before deployment, identifying the groups likely to be affected, the risks they may face~\cite[Article~27]{europeanparliamentandcounciloftheeuropeanunionRegulationEU20242024}. Both frameworks recognize that the consequences of an AI system depend on more than its technical properties. They also establish mechanisms for examining those consequences in context and, in the case of the NIST framework, across the system's life cycle. But neither, by itself, resolves the question of whose account of that context should prevail. An impact assessment can document affected groups without giving them influence over the decisions that follow. A risk-management process can monitor a system without questioning the institutional purposes it serves. The existence of a requirement to assess impacts does not determine what counts as an acceptable impact, or who has the authority to make that judgment. These frameworks therefore offer institutional footholds for thick alignment, but their significance depends on how they are implemented and who is able to participate in that implementation.

The third emphasis raises a more difficult question for policy: what would it mean for affected publics to have a part in deciding which values guide AI systems? The philosopher John Dewey described a public as those who are affected by the indirect consequences of transactions they did not enter~\cite{deweyPublicItsProblems1927a}. AI systems can create publics in this sense: people whose opportunities, treatment, or access to services are shaped by systems they had no part in designing and may have little ability to contest. Soliciting their feedback is not the same as giving them a role in determining how those systems should operate. Nor does a representative sample consulted by a private company necessarily provide the democratic authorization that public decision-making requires. The Blueprint for an AI Bill of Rights, issued by the White House Office of Science and Technology Policy in 2022, called for public consultation and direct engagement with diverse impacted communities throughout the design, development, and deployment of automated systems~\cite{whitehouseofficeofscienceandtechnologypolicyBlueprintAIBill2022}. Such principles point toward a broader role for affected publics that stops short of binding decision-making authority. Other democratic institutions offer resources for thinking about what that authority might involve. Citizens' assemblies, deliberative bodies of randomly selected members of the public that Hélène Landemore~\cite{landemore2020open} sees as a way to broaden political inclusion, have informed constitutional reform in Ireland and proposals for climate policy in France~\cite{farrellReimaginingDemocracyLessons2019,giraudetCoconstructionDeliberativeDemocracy2022}. These experiences do not provide a ready-made model for AI governance, but they demonstrate that public deliberation can be organized as something more substantial than the collection of individual preferences. Applied to AI, such arrangements could give affected communities a role in examining the purposes of a system, the trade-offs it entails, and the conditions under which its use should be permitted. They would also require institutions to explain how public deliberations inform decisions and what avenues remain available when those decisions are contested.

This does not mean that every decision about AI should be made locally, or that common standards have no place in governance. Walzer's distinction between thin and thick, introduced above, offers a way to think about how shared principles and situated judgments might coexist. A thin layer of agreement can establish terms that people across different societies recognize, while the thicker meanings of those principles are worked out in particular social and political settings. Policy can play a role at both levels. Common requirements for transparency, documentation, assessment, and accountability can make systems more legible and provide grounds on which their development and use can be challenged. But these requirements cannot settle in advance what fairness means in a particular hiring process, what constitutes acceptable risk in a system used by a court, or whose interests should prevail when the priorities of patients and clinicians diverge. Those judgments require institutions capable of attending to local circumstances and sustaining disagreement about the values at stake.

The challenge for AI governance is not to choose between common standards and contextual judgment; it is to establish arrangements through which they can be mutually informing rather than displacing. Thick alignment asks policymakers to enable and sustain the conditions under which values can be described, examined, and contested, and under which those affected by AI systems have a meaningful part in deciding how they are used.

\section{Conclusion}

AI alignment research has made human values a technical problem, something to be specified and optimized. We have argued that this approach rests on a limited account of the relationship between technology and values, of what values are, and how they acquire meaning. Values are interpreted in social contexts, constituted through practices and institutions, and contested among people whose interests and authority are unequally distributed. Thin alignment obscures these processes when it converts situated judgments into a small number of measures or metrics. Thick alignment offers a different starting point. It asks how values take shape in the settings where AI systems are developed and used, how they change across the life cycle of those systems, where people disagree about them, and whose judgments determine what the systems should do.

Consider again the GPT-4o update with which we began. Its effectiveness was judged against feedback from raters and users. Its success was understood in terms of how well the model responded to their preferences. But responsiveness to feedback is not the same as understanding what that feedback means. A thumbs-up registers approval without distinguishing whether a user is expressing agreement, acknowledging a response, or simply signaling a desire to move on. Nor does it tell us whether the response that pleased one user served that person's interests, or whether it might have encouraged beliefs or plans that harmed them~\cite{hill2025confided}. The product update raises issues that could not be answered by measuring alignment with preference labeling. Whether GPT-4o had been aligned successfully depended on whose expectations shaped the model and on how these expectations came to stand for human preferences more broadly.

Thick alignment does not supply a more comprehensive specification of human values, nor does it promise to resolve the disagreements that arise when different values and interests meet. It asks instead that alignment research take those disagreements seriously and attend to the practices through which they are negotiated. For researchers, this means studying systems in the settings where they are used, following them from development into use, and engaging the publics they affect as integral parts and partners of research and evaluation rather than as additions to technical assessment. For policymakers, it means building arrangements that make value choices visible, preserve opportunities to contest them, and give people meaningful influence over decisions. These are not preliminaries to action; they are the actions through which alignment is done. Thick alignment seeks to keep the judgments built into these systems, and the consequences that follow, open to public scrutiny and democratic challenge.



\begin{acks}
This research was supported by the Ford Foundation and the Heising-Simons Foundation. We thank Christine Custis and Nicholas Collins for their support of this work, and the members of the Science, Technology \& Social Values Lab at the Institute for Advanced Study---especially Carina Albrecht, Sareeta Amrute, Brian Jordan Jefferson, Inioluwa Deborah Raji, and Alejandra Ruiz Leon---for their generous engagement with the ideas developed here. We are grateful to the audiences at the 2023 ACM Conference on Fairness, Accountability, and Transparency and the 2023 Institute for Ethics in AI Annual Lecture at the University of Oxford, whose questions helped to refine these arguments.
\end{acks}

\section*{Ethics and Privacy Statement}
Our conceptual analysis uses only published scholarship and public documents and involves no human participants, new data, or new system. We anticipate minimal societal risks beyond the risks the paper examines.

\section*{Generative AI Disclosure Statement}

Claude 3 Opus was used for copyediting earlier drafts of this paper. The authors reviewed all suggested edits. ChatGPT (GPT-6 Astra; accessed September 2026) was used to support citation tracing and broader literature discovery concerning the concept of ``thick alignment.'' This use supplemented conventional searches conducted through Google Scholar to identify related research areas and potentially relevant grey literature beyond the sources found in those searches. All authors retain full responsibility for the originality, accuracy, and integrity of the manuscript, including the correctness, authenticity, and appropriate attribution of all references and citations.

\bibliographystyle{ACM-Reference-Format}
\bibliography{draft_v3_1}

@article{albrechtDiscerningAudiencesButtons2024,
  title = {Discerning {{Audiences Through Like Buttons}}},
  author = {Albrecht, Carina},
  year = 2024,
  month = jan,
  journal = {Harvard Data Science Review},
  volume = {6},
  number = {1},
  publisher = {The MIT Press},
  issn = {2644-2353, 2688-8513},
  doi = {10.1162/99608f92.e3d09bff},
  urldate = {2026-11-03},
  langid = {english}
}

@misc{alikhaniHypeHarmWhy2025,
  title = {Hype and Harm: {{Why}} We Must Ask Harder Questions about {{AI}} and Its Alignment with Human Values},
  shorttitle = {Hype and Harm},
  author = {Alikhani, Malihe and Hassan, Sabit},
  year = 2025,
  month = oct,
  urldate = {2026-09-29},
  howpublished = {Brookings},
  langid = {american},
  url = {https://www.brookings.edu/articles/hype-and-harm-why-we-must-ask-harder-questions-about-ai-and-its-alignment-with-human-values/}
}

@article{angwinMachineBias2016,
  title = {Machine {{Bias}}},
  author = {Angwin, Julia and Larson, Jeff and Mattu, Surya and Kirchner, Lauren},
  year = 2016,
  month = may,
  journal = {ProPublica},
  langid = {american},
  note = {\url{https://www.propublica.org/article/machine-bias-risk-assessments-in-criminal-sentencing}}
}

@misc{anthropicClaudesConstitution2026,
  title = {Claude's Constitution},
  author = {{Anthropic}},
  year = 2026,
  month = jan,
  howpublished = {Anthropic},
  url = {https://www.anthropic.com/constitution}
}

@misc{baiConstitutionalAIHarmlessness2022,
  title = {Constitutional {{AI}}: {{Harmlessness}} from {{AI Feedback}}},
  shorttitle = {Constitutional {{AI}}},
  author = {Bai, Yuntao and Kadavath, Saurav and Kundu, Sandipan and Askell, Amanda and Kernion, Jackson and Jones, Andy and Chen, Anna and Goldie, Anna and Mirhoseini, Azalia and McKinnon, Cameron and Chen, Carol and Olsson, Catherine and Olah, Christopher and Hernandez, Danny and Drain, Dawn and Ganguli, Deep and Li, Dustin and {Tran-Johnson}, Eli and Perez, Ethan and Kerr, Jamie and Mueller, Jared and Ladish, Jeffrey and Landau, Joshua and Ndousse, Kamal and Lukosuite, Kamile and Lovitt, Liane and Sellitto, Michael and Elhage, Nelson and Schiefer, Nicholas and Mercado, Noemi and DasSarma, Nova and Lasenby, Robert and Larson, Robin and Ringer, Sam and Johnston, Scott and Kravec, Shauna and Showk, Sheer El and Fort, Stanislav and Lanham, Tamera and {Telleen-Lawton}, Timothy and Conerly, Tom and Henighan, Tom and Hume, Tristan and Bowman, Samuel R. and {Hatfield-Dodds}, Zac and Mann, Ben and Amodei, Dario and Joseph, Nicholas and McCandlish, Sam and Brown, Tom and Kaplan, Jared},
  year = 2022,
  month = dec,
  number = {arXiv:2212.08073},
  eprint = {2212.08073},
  primaryclass = {cs.CL},
  publisher = {arXiv},
  doi = {10.48550/arXiv.2212.08073},
  urldate = {2026-10-06},
  archiveprefix = {arXiv}
}

@book{benjaminRaceTechnologyAbolitionist2019,
  title = {Race after Technology: {{Abolitionist}} Tools for the {{New Jim Code}}},
  author = {Benjamin, Ruha},
  year = 2019,
  publisher = {Polity Press},
  address = {Cambridge, UK}
}

@misc{bogenAssessingAISurveying2026,
  title = {Assessing {{AI}}: {{Surveying}} the {{Spectrum}} of {{Approaches}} to {{Understanding}} and {{Auditing AI Systems}}},
  shorttitle = {Assessing {{AI}}},
  author = {Bogen, Miranda},
  year = 2026,
  month = jan,
  urldate = {2026-09-29},
  howpublished = {Center for Democracy and Technology},
  langid = {american},
  url = {https://cdt.org/insights/assessing-ai-surveying-the-spectrum-of-approaches-to-understanding-and-auditing-ai-systems/}
}

@book{boltanskiJustificationEconomiesWorth2006,
  title = {On {{Justification}}: {{Economies}} of {{Worth}}},
  shorttitle = {On {{Justification}}},
  author = {Boltanski, Luc and Th{\'e}venot, Laurent},
  year = 2006,
  publisher = {Princeton University Press},
  address = {Princeton, NJ},
  googlebooks = {9DlFDwAAQBAJ},
  isbn = {978-0-691-12516-9},
  langid = {english}
}

@article{brandtEngagingAdvisoryBoard2025,
  title = {Engaging an Advisory Board in Discussions about the Ethical Relevance of Algorithmic Bias and Fairness},
  author = {Brandt, Laura and Au, Larry and Castro, Clinton and Odom, Gabriel J.},
  year = 2025,
  month = may,
  journal = {npj Digital Medicine},
  volume = {8},
  number = {1},
  pages = {292},
  publisher = {Nature Publishing Group},
  issn = {2398-6352},
  doi = {10.1038/s41746-025-01711-1},
  urldate = {2026-09-29},
  copyright = {2025 The Author(s)},
  langid = {english}
}

@inproceedings{buolamwiniGenderShadesIntersectional2018,
  title = {Gender {{Shades}}: {{Intersectional Accuracy Disparities}} in {{Commercial Gender Classification}}},
  shorttitle = {Gender {{Shades}}},
  booktitle = {Proceedings of the 1st {{Conference}} on {{Fairness}}, {{Accountability}} and {{Transparency}}},
  author = {Buolamwini, Joy and Gebru, Timnit},
  year = 2018,
  month = jan,
  pages = {77--91},
  publisher = {PMLR},
  issn = {2640-3498},
  urldate = {2026-10-07},
  langid = {english},
  note = {\url{https://proceedings.mlr.press/v81/buolamwini18a.html}}
}

@article{chouldechovaFairPredictionDisparate2017,
  title = {Fair {{Prediction}} with {{Disparate Impact}}: {{A Study}} of {{Bias}} in {{Recidivism Prediction Instruments}}},
  shorttitle = {Fair {{Prediction}} with {{Disparate Impact}}},
  author = {Chouldechova, Alexandra},
  year = 2017,
  month = jun,
  journal = {Big Data},
  volume = {5},
  number = {2},
  pages = {153--163},
  publisher = {SAGE Publications},
  issn = {2167-6461},
  doi = {10.1089/big.2016.0047},
  urldate = {2026-10-07},
  langid = {english}
}

@book{christianAlignmentProblemMachine2020,
  title = {The {{Alignment Problem}}: {{Machine Learning}} and {{Human Values}}},
  shorttitle = {The {{Alignment Problem}}},
  author = {Christian, Brian},
  year = 2020,
  month = oct,
  publisher = {W. W. Norton \& Company},
  googlebooks = {KGCNEAAAQBAJ},
  isbn = {978-0-393-63582-9},
  langid = {english},
  address = {New York, NY}
}

@inproceedings{christianoDeepReinforcementLearning2017,
  title = {Deep {{Reinforcement Learning}} from {{Human Preferences}}},
  booktitle = {Advances in {{Neural Information Processing Systems}}},
  author = {Christiano, Paul F. and Leike, Jan and Brown, Tom and Martic, Miljan and Legg, Shane and Amodei, Dario},
  year = 2017,
  volume = {30},
  publisher = {Curran Associates, Inc.},
  urldate = {2026-10-07},
  pages = {4299--4307},
  note = {\url{https://proceedings.neurips.cc/paper_files/paper/2017/hash/d5e2c0adad503c91f91df240d0cd4e49-Abstract.html}}
}

@article{cuthbertson2025chatgpt,
  title = {{{ChatGPT}} Is Pushing People towards Mania, Psychosis and Death -- and {{OpenAI}} Doesn't Know How to Stop It},
  author = {Cuthbertson, Anthony},
  year = 2025,
  month = jul,
  journal = {The Independent},
  note = {\url{https://www.the-independent.com/tech/chatgpt-ai-therapy-chatbot-psychosis-mental-health-b2797487.html}}
}

@article{dastin2018amazon,
  title = {Amazon Scraps Secret {{AI}} Recruiting Tool That Showed Bias against Women},
  author = {Dastin, Jeffrey},
  year = 2018,
  month = oct,
  journal = {Reuters},
  note = {\url{https://www.reuters.com/article/us-amazon-com-jobs-automation-insight/amazon-scraps-secret-ai-recruiting-tool-that-showed-bias-against-women-idUSKCN1MK08G/}}
}

@book{deweyPublicItsProblems1927a,
  title = {The Public and Its Problems},
  author = {Dewey, John},
  year = 1927,
  publisher = {{Henry Holt and Company}},
  address = {New York, NY}
}

@article{espelandCommensurationSocialProcess1998,
  title = {Commensuration as a {{Social Process}}},
  author = {Espeland, Wendy Nelson and Stevens, Mitchell L.},
  year = 1998,
  journal = {Annual Review of Sociology},
  volume = {24},
  pages = {313--343},
  publisher = {Annual Reviews},
  issn = {1545-2115},
  doi = {10.1146/annurev.soc.24.1.313},
  urldate = {2026-10-07}
}

@misc{europeanparliamentandcounciloftheeuropeanunionRegulationEU20242024,
  title = {Regulation ({{EU}}) 2024/1689 of the {{European Parliament}} and of the {{Council}} of 13 {{June}} 2024 Laying down Harmonised Rules on Artificial Intelligence ({{Artificial Intelligence Act}})},
  author = {{European Parliament and Council of the European Union}},
  year = 2024,
  howpublished = {Official Journal of the European Union, L, 2024/1689, 12 July 2024}
}

@book{farrellReimaginingDemocracyLessons2019,
  title = {Reimagining Democracy: {{Lessons}} in Deliberative Democracy from the {{Irish}} Front Line},
  author = {Farrell, David M. and Suiter, Jane},
  year = 2019,
  publisher = {Cornell University Press},
  address = {Ithaca, NY}
}

@inproceedings{fleisigPerspectivistParadigmShift2024,
  title = {The {{Perspectivist Paradigm Shift}}: {{Assumptions}} and {{Challenges}} of {{Capturing Human Labels}}},
  shorttitle = {The {{Perspectivist Paradigm Shift}}},
  booktitle = {Proceedings of the 2024 {{Conference}} of the {{North American Chapter}} of the {{Association}} for {{Computational Linguistics}}: {{Human Language Technologies}} ({{Volume}} 1: {{Long Papers}})},
  author = {Fleisig, Eve and Blodgett, Su Lin and Klein, Dan and Talat, Zeerak},
  editor = {Duh, Kevin and Gomez, Helena and Bethard, Steven},
  year = 2024,
  month = jun,
  pages = {2279--2292},
  publisher = {Association for Computational Linguistics},
  address = {Mexico City, Mexico},
  doi = {10.18653/v1/2024.naacl-long.126},
  urldate = {2026-11-04}
}

@article{fosterThinThickPolitics2023,
  title = {From Thin to Thick: {{Toward}} a Politics of Human-Compatible {{AI}}},
  author = {Foster, Jacob G.},
  year = 2023,
  journal = {Public Culture},
  volume = {35},
  number = {3},
  pages = {417--430},
  doi = {10.1215/08992363-10742593}
}

@article{gebruDatasheetsDatasets2021,
  title = {Datasheets for Datasets},
  author = {Gebru, Timnit and Morgenstern, Jamie and Vecchione, Briana and Vaughan, Jennifer Wortman and Wallach, Hanna and III, Hal Daum{\'e} and Crawford, Kate},
  year = 2021,
  month = nov,
  journal = {Communications of the ACM},
  volume = {64},
  number = {12},
  pages = {86--92},
  issn = {0001-0782},
  doi = {10.1145/3458723},
  urldate = {2026-10-07}
}

@incollection{geertz1973thick,
  title = {Thick Description: {{Toward}} an Interpretive Theory of Culture},
  booktitle = {The Interpretation of Cultures: {{Selected}} Essays},
  author = {Geertz, Clifford},
  year = 1973,
  pages = {3--30},
  publisher = {Basic Books},
  address = {New York}
}

@article{gerlitzEconomySocialButtons2013,
  title = {The {{Like Economy}}: {{Social Buttons}} and the {{Data-Intensive Web}}},
  shorttitle = {The {{Like Economy}}},
  author = {Gerlitz, Carolin and Helmond, Anne},
  year = 2013,
  month = dec,
  journal = {New Media \& Society},
  volume = {15},
  number = {8},
  pages = {1348--1365},
  publisher = {SAGE Publications},
  issn = {1461-4448},
  doi = {10.1177/1461444812472322},
  urldate = {2026-10-07},
  langid = {english}
}

@article{giraudetCoconstructionDeliberativeDemocracy2022,
  title = {``{{Co}}-Construction'' in Deliberative Democracy: {{Lessons}} from the {{French Citizens}}' {{Convention}} for {{Climate}}},
  author = {Giraudet, Louis-Ga{\"e}tan and Apouey, B{\'e}n{\'e}dicte and Arab, Hazem and Baeckelandt, Simon and B{\'e}gout, Philippe and Berghmans, Nicolas and Blanc, Nathalie and others},
  year = 2022,
  journal = {Humanities and Social Sciences Communications},
  volume = {9},
  number = {1},
  pages = {207},
  doi = {10.1057/s41599-022-01212-6}
}

@inproceedings{greenFalsePromiseRisk2020,
  title = {The False Promise of Risk Assessments: Epistemic Reform and the Limits of Fairness},
  shorttitle = {The False Promise of Risk Assessments},
  booktitle = {Proceedings of the 2020 {{Conference}} on {{Fairness}}, {{Accountability}}, and {{Transparency}}},
  author = {Green, Ben},
  year = 2020,
  month = jan,
  series = {{{FAT}}* '20},
  pages = {594--606},
  publisher = {Association for Computing Machinery},
  address = {New York, NY, USA},
  doi = {10.1145/3351095.3372869},
  urldate = {2026-11-03},
  isbn = {978-1-4503-6936-7}
}

@misc{hammondMultiAgentRisksAdvanced2025,
  title = {Multi-{{Agent Risks}} from {{Advanced AI}}},
  author = {Hammond, Lewis and Chan, Alan and Clifton, Jesse and {Hoelscher-Obermaier}, Jason and Khan, Akbir and McLean, Euan and Smith, Chandler and Barfuss, Wolfram and Foerster, Jakob and Gaven{\v c}iak, Tom{\'a}{\v s} and Han, The Anh and Hughes, Edward and Kova{\v r}{\'i}k, Vojt{\v e}ch and Kulveit, Jan and Leibo, Joel Z. and Oesterheld, Caspar and de Witt, Christian Schroeder and Shah, Nisarg and Wellman, Michael and Bova, Paolo and Cimpeanu, Theodor and Ezell, Carson and {Feuillade-Montixi}, Quentin and Franklin, Matija and Kran, Esben and Krawczuk, Igor and Lamparth, Max and Lauffer, Niklas and Meinke, Alexander and Motwani, Sumeet and Reuel, Anka and Conitzer, Vincent and Dennis, Michael and Gabriel, Iason and Gleave, Adam and Hadfield, Gillian and Haghtalab, Nika and Kasirzadeh, Atoosa and Krier, S{\'e}bastien and Larson, Kate and Lehman, Joel and Parkes, David C. and Piliouras, Georgios and Rahwan, Iyad},
  year = 2025,
  month = feb,
  number = {arXiv:2502.14143},
  eprint = {2502.14143},
  primaryclass = {cs.MA},
  publisher = {arXiv},
  doi = {10.48550/arXiv.2502.14143},
  urldate = {2026-09-29},
  archiveprefix = {arXiv}
}

@misc{henselmansThinAlignmentCant2025,
  title = {Thin {{Alignment Can}}'t {{Solve Thick Problems}}},
  author = {Henselmans, Daan},
  year = 2025,
  month = apr,
  urldate = {2026-09-29},
  howpublished = {LessWrong},
  url = {https://www.lesswrong.com/posts/uyfmdMKGjYHqiLbbX/thin-alignment-can-t-solve-thick-problems}
}

@article{hill2025confided,
  title = {A Teen Was Suicidal. {{ChatGPT}} Was the Friend He Confided In},
  author = {Hill, Kashmir},
  year = 2025,
  month = aug,
  journal = {The New York Times},
  note = {\url{https://www.nytimes.com/2025/08/26/technology/chatgpt-openai-suicide.html}}
}

@article{hillLawsuitsBlameChatGPT2025,
  title = {Lawsuits Blame {{ChatGPT}} for Suicides and Harmful Delusions},
  author = {Hill, Kashmir},
  year = 2025,
  month = nov,
  journal = {The New York Times},
  note = {\url{https://www.nytimes.com/2025/11/06/technology/chatgpt-lawsuit-suicides-delusions.html}}
}

@misc{huggingface2026security,
  title = {Security Incident Disclosure---{{July}} 2026},
  author = {{Hugging Face}},
  year = 2026,
  month = jul,
  howpublished = {Hugging Face blog},
  url = {https://huggingface.co/blog/security-incident-july-2026}
}

@book{jackson2013thin,
  title = {Thin Description: {{Ethnography}} and the {{African Hebrew Israelites}} of {{Jerusalem}}},
  author = {Jackson, Jr., John L.},
  year = 2013,
  publisher = {Harvard University Press},
  address = {Cambridge, MA}
}

@book{jasanoffStatesKnowledgeCoproduction2004,
  title = {States of {{Knowledge}}: {{The Co}}-Production of {{Science}} and {{Social Order}}},
  shorttitle = {States of {{Knowledge}}},
  editor = {Jasanoff, Sheila},
  year = 2004,
  publisher = {Routledge},
  address = {London},
  googlebooks = {zYTgAAAAMAAJ},
  isbn = {978-0-415-33361-0},
  langid = {english}
}

@misc{jhaHumanAmplificationIntelligent2026,
  title = {Human Amplification, Intelligent Agents, and the Aim of {{AI}} Research},
  author = {Jha, Tushita and Everitt, Tom and Grzankowski, Alex},
  year = 2026,
  month = feb,
  howpublished = {PhilArchive preprint},
  url = {https://philarchive.org/rec/JHAHAI}
}

@inproceedings{jhaPositionAIAlignment2026,
  title = {Position: ``{{AI Alignment}}'' {{Encompasses Competing Technical Priorities}}},
  booktitle = {{{ICML}} 2026},
  author = {Jha, Tushita and Svarc, Rory and Bagi{\'n}ski, Mateusz},
  year = 2026,
  langid = {english}
}

@book{keaneAnimalsRobotsGods2025,
  title = {Animals, {{Robots}}, {{Gods}}: {{Adventures}} in the {{Moral Imagination}}},
  shorttitle = {Animals, {{Robots}}, {{Gods}}},
  author = {Keane, Webb},
  year = 2025,
  month = feb,
  publisher = {Princeton University Press},
  googlebooks = {cXchEQAAQBAJ},
  isbn = {978-0-691-27093-7},
  langid = {english},
  address = {Princeton, NJ}
}

@inproceedings{kirkPRISMAlignmentDataset2024,
  title = {The {{PRISM}} Alignment Dataset: What Participatory, Representative and Individualised Human Feedback Reveals about the Subjective and Multicultural Alignment of Large Language Models},
  shorttitle = {The {{PRISM}} Alignment Dataset},
  booktitle = {Proceedings of the 38th {{International Conference}} on {{Neural Information Processing Systems}}},
  author = {Kirk, Hannah Rose and Whitefield, Alexander and R{\"o}ttger, Paul and Bean, Andrew and Margatina, Katerina and Ciro, Juan and Mosquera, Rafael and Bartolo, Max and Williams, Adina and He, He and Vidgen, Bertie and Hale, Scott A.},
  year = 2024,
  month = dec,
  series = {{{NeurIPS}} '24},
  volume = {37},
  pages = {105236--105344},
  publisher = {Curran Associates Inc.},
  address = {Red Hook, NY, USA},
  urldate = {2026-10-06},
  isbn = {979-8-3313-1438-5}
}

@article{kirkWhyHumanAI2025,
  title = {Why Human--{{AI}} Relationships Need Socioaffective Alignment},
  author = {Kirk, Hannah Rose and Gabriel, Iason and Summerfield, Chris and Vidgen, Bertie and Hale, Scott A.},
  year = 2025,
  month = may,
  journal = {Humanities and Social Sciences Communications},
  volume = {12},
  number = {1},
  pages = {728},
  publisher = {Palgrave},
  issn = {2662-9992},
  doi = {10.1057/s41599-025-04532-5},
  urldate = {2026-09-29},
  copyright = {2025 Crown},
  langid = {english}
}

@inproceedings{kleinberg2017inherent,
  title = {Inherent Trade-Offs in the Fair Determination of Risk Scores},
  booktitle = {8th {{Innovations}} in {{Theoretical Computer Science Conference}} ({{ITCS}} 2017)},
  author = {Kleinberg, Jon and Mullainathan, Sendhil and Raghavan, Manish},
  year = 2017,
  series = {Leibniz International Proceedings in Informatics (Lipics)},
  publisher = {Schloss Dagstuhl--Leibniz-Zentrum für Informatik},
  volume = {67},
  pages = {43:1--43:23},
  doi = {10.4230/LIPIcs.ITCS.2017.43}
}

@book{lamont2009professors,
  title = {How Professors Think: {{Inside}} the Curious World of Academic Judgment},
  author = {Lamont, Mich{\`e}le},
  year = 2009,
  publisher = {Harvard University Press},
  address = {Cambridge, MA}
}

@article{lamontComparativeSociologyValuation2012,
  title = {Toward a {{Comparative Sociology}} of {{Valuation}} and {{Evaluation}}},
  author = {Lamont, Mich{\`e}le},
  year = 2012,
  journal = {Annual Review of Sociology},
  volume = {38},
  pages = {201--221},
  publisher = {Annual Reviews},
  issn = {1545-2115},
  doi = {10.1146/annurev-soc-070308-120022},
  urldate = {2026-10-07}
}

@book{landemore2020open,
  title = {Open Democracy: {{Reinventing}} Popular Rule for the Twenty-First Century},
  author = {Landemore, H{\'e}l{\`e}ne},
  year = 2020,
  publisher = {Princeton University Press},
  address = {Princeton, NJ},
  isbn = {978-0-691-18199-8}
}

@article{loveCloseReadingThin2013,
  title = {Close {{Reading}} and {{Thin Description}}},
  author = {Love, Heather},
  year = 2013,
  month = sep,
  journal = {Public Culture},
  volume = {25},
  number = {3 (71)},
  pages = {401--434},
  issn = {0899-2363},
  doi = {10.1215/08992363-2144688},
  urldate = {2026-10-07}
}

@inproceedings{mitchellModelCardsModel2019,
  title = {Model {{Cards}} for {{Model Reporting}}},
  booktitle = {Proceedings of the {{Conference}} on {{Fairness}}, {{Accountability}}, and {{Transparency}}},
  author = {Mitchell, Margaret and Wu, Simone and Zaldivar, Andrew and Barnes, Parker and Vasserman, Lucy and Hutchinson, Ben and Spitzer, Elena and Raji, Inioluwa Deborah and Gebru, Timnit},
  year = 2019,
  month = jan,
  series = {{{FAT}}* '19},
  pages = {220--229},
  publisher = {Association for Computing Machinery},
  address = {New York, NY, USA},
  doi = {10.1145/3287560.3287596},
  urldate = {2026-10-06},
  isbn = {978-1-4503-6125-5}
}

@inproceedings{movvaWhatsMyHuman2025,
  title = {What's {{In My Human Feedback}}? {{Learning Interpretable Descriptions}} of {{Preference Data}}},
  shorttitle = {What's {{In My Human Feedback}}?},
  booktitle = {The {{Fourteenth International Conference}} on {{Learning Representations}}},
  author = {Movva, Rajiv and Milli, Smitha and Min, Sewon and Pierson, Emma},
  year = 2025,
  month = oct,
  urldate = {2026-11-04},
  langid = {english},
  note = {\url{https://openreview.net/forum?id=sC6A1bFDUt}}
}

@techreport{nationalinstituteofstandardsandtechnologyArtificialIntelligenceRisk2023,
  type = {{{NIST AI}}},
  title = {Artificial Intelligence Risk Management Framework ({{AI RMF}} 1.0)},
  author = {{National Institute of Standards and Technology}},
  year = 2023,
  number = {100-1},
  address = {Gaithersburg, MD},
  institution = {U.S. Department of Commerce},
  doi = {10.6028/NIST.AI.100-1}
}

@misc{nelsonThickAlignment2023,
  title = {Thick {{Alignment}}},
  author = {Nelson, Alondra},
  year = 2023,
  month = oct,
  howpublished = {Annual Lecture, Institute for Ethics in AI, University of Oxford, 19 October 2023. Video},
  url = {https://www.youtube.com/watch?v=7FaWxNwBVTM}
}

@misc{nelsonThickAlignment2023a,
  title = {Thick {{Alignment}}},
  author = {Nelson, Alondra},
  year = 2023,
  month = jun,
  howpublished = {Opening keynote address, ACM Conference on Fairness, Accountability, and Transparency (FAccT 2023), Chicago, 12 June 2023. Video},
  url = {https://www.youtube.com/watch?v=Sq_XwqVTqvQ}
}

@misc{openaiSycophancyGPT4oWhat2025,
  title = {Sycophancy in {{GPT-4}}o: {{What}} Happened and What We're Doing about It},
  shorttitle = {Sycophancy in {{GPT-4}}o},
  author = {{OpenAI}},
  year = 2025,
  month = apr,
  urldate = {2026-10-07},
  howpublished = {OpenAI},
  langid = {american},
  url = {https://openai.com/index/sycophancy-in-gpt-4o/}
}

@inproceedings{ouyangTrainingLanguageModels2022,
  title = {Training Language Models to Follow Instructions with Human Feedback},
  booktitle = {Proceedings of the 36th {{International Conference}} on {{Neural Information Processing Systems}}},
  author = {Ouyang, Long and Wu, Jeff and Jiang, Xu and Almeida, Diogo and Wainwright, Carroll L. and Mishkin, Pamela and Zhang, Chong and Agarwal, Sandhini and Slama, Katarina and Ray, Alex and Schulman, John and Hilton, Jacob and Kelton, Fraser and Miller, Luke and Simens, Maddie and Askell, Amanda and Welinder, Peter and Christiano, Paul and Leike, Jan and Lowe, Ryan},
  year = 2022,
  month = nov,
  series = {{{NeurIPS}} '22},
  pages = {27730--27744},
  publisher = {Curran Associates Inc.},
  address = {Red Hook, NY, USA},
  urldate = {2026-10-06},
  isbn = {978-1-7138-7108-8}
}

@article{quillianMetaanalysisFieldExperiments2017,
  title = {Meta-Analysis of Field Experiments Shows No Change in Racial Discrimination in Hiring over Time},
  author = {Quillian, Lincoln and Pager, Devah and Hexel, Ole and Midtb{\o}en, Arnfinn H.},
  year = 2017,
  month = oct,
  journal = {Proceedings of the National Academy of Sciences},
  volume = {114},
  number = {41},
  pages = {10870--10875},
  publisher = {Proceedings of the National Academy of Sciences},
  doi = {10.1073/pnas.1706255114},
  urldate = {2026-11-02}
}

@incollection{ryleThinkingThoughtsWhat1971,
  title = {The {{Thinking}} of {{Thoughts}}: {{What Is}} ``{{Le Penseur}}'' {{Doing}}?},
  booktitle = {Collected {{Papers}}},
  author = {Ryle, Gilbert},
  year = 1971,
  volume = {2},
  publisher = {Hutchinson},
  address = {London},
  origlocation = {University of Saskatchewan},
  origpublisher = {University Lectures no. 18},
  pages = {480--496}
}

@inproceedings{santurkarWhoseOpinionsLanguage2023,
  title = {Whose {{Opinions Do Language Models Reflect}}?},
  booktitle = {Proceedings of the 40th {{International Conference}} on {{Machine Learning}}},
  author = {Santurkar, Shibani and Durmus, Esin and Ladhak, Faisal and Lee, Cinoo and Liang, Percy and Hashimoto, Tatsunori},
  year = 2023,
  month = jul,
  series = {{{ICML}}'23},
  volume = {202},
  pages = {29971--30004},
  publisher = {JMLR.org},
  address = {Honolulu, Hawaii, USA},
  urldate = {2026-10-06}
}

@article{schmer-galunderAnthropologyAI2025,
  title = {Anthropology for {{AI}}},
  author = {{Schmer-Galunder}, Sonja},
  year = 2025,
  month = mar,
  journal = {Collegium Antropologicum},
  volume = {49},
  number = {1},
  pages = {1--6},
  issn = {1848-9486},
  urldate = {2026-09-29},
  copyright = {All authors submitting their article for publication have to agree to the  Authorship statement .     IMPORTANT:    Authorship statement form   should be dowloaded from    HERE   , printed out, filled out,   signed by ALL CO-AUTHORS   and then sent in   regular mail   to Collegium Antropologicum journal office (address: Collegium Antropologicum, HAD, Hebrangova 1, HR-10000 Zagreb, Croatia). In the form, the co-authors must be listed in the order they would appear on the article, should the manuscript be published. The order of the coauthors listed on the Authorship statement form must correspond to the order of authors entered in the  Step 3  of the online submission procedure ( Entering the Submission's Metadata ).   All subsequent changes in the order or the number of authors   must be accompanied by a new Authorship Statement form signed again by all co-authors.    Please note that the manuscript will not be processed until journal office receives a valid Authorship statement form!          Authorship statement:     1. Copyright issues    The listed authors warrant that they are the authors and sole owners of the submitted manuscript. The authors also warrant that the work is original; that it has not been previously published in print or electronic format and is not under consideration by another publisher or electronic medium; that it has not been previously transferred, assigned, or otherwise encumbered; and that the authors have full power to grant such rights. With respect to the results of this work, the manuscript of this or substantially similar content will not be submitted to any other journal until the review process in the Collegium Antropologicum has been officially completed (acceptance or rejection of the manuscript).    The paper will not be withdrawn from the review process by the Collegium Antropologicum Editorial Board until the review process is completed. The authors will comply with the requests of the Collegium Antropologicum Editors and reviewers to improve the paper for publication. The eventual disagreements will be submitted in a written form; the authors are aware that the disagreement(s) with the Collegium Antropologicum's requests may result in the rejection of the manuscript. The authors hereby grant to the Collegium Antropologicum the right to edit, revise, abridge, and condense the manuscript.    If the manuscript is accepted for publication in the Collegium Antropologicum, the authors hereby transfer the copyright of the paper to the Collegium Antropologicum. The authors permit the Collegium Antropologicum to allow third parties to copy any part of the journal without asking for permission, provided that the reference to the source is given.    For papers with more than one author: All other co-authors agree to allow the corresponding author to make decisions regarding prepublication release of the information in the paper to the media, government agencies, or both.     2. Financial disclosure and conflict of interest      Authors certify that all financial and material support for this research and work is clearly identified in the manuscript.    Authors certify that all affiliations with or financial involvement (e.g. employment, consultancies, honoraria, stock ownership or options, expert testimony, grants or patents received or pending, royalties) with any organization or entity with a financial interest in or financial conflict with the subject or materials discussed in the manuscript, along with all other financial~and personal relationships that might bias the authors' work   will be disclosed to the Editorial Board as part of submission .       3. Ethical standards      Authors confirm that the procedures followed in the manuscript were in accordance with the ethical standards of the responsible institution in the community or organization that scientists work in.     4. Permissions      Authors certify that copies of all permissions to reproduce published material, to use illustrations or report information about identifiable people, or to name people for their contributions   are submitted along with the manuscript  .},
  langid = {english},
  doi = {10.5671/ca.49.1.1}
}

@article{schmer-galunderCultureCodeAnthropological2026,
  title = {Culture in the Code: {{Anthropological Concepts Decoding AI}}'s {{Hidden Assumptions}}},
  shorttitle = {Culture in the Code},
  author = {{Schmer-Galunder}, Sonja},
  year = 2026,
  month = apr,
  journal = {AI and Ethics},
  volume = {6},
  number = {3},
  pages = {272},
  issn = {2730-5961},
  doi = {10.1007/s43681-026-01122-2},
  urldate = {2026-09-29},
  langid = {english}
}

@techreport{schwartzAssessingRisksImpacts2024,
  title = {The {{Assessing Risks}} and {{Impacts}} of {{AI}} ({{ARIA}}) {{Program Evaluation Design Document}}},
  author = {Schwartz, Reva and Waters, Gabriella and Amironesei, Razvan and Fiscus, Jon and Hall, Patrick and Jones, Anya and Jain, Shomik and Godil, Afzal and Greene, Kristen and Jensen, Ted and Schulman, Noah},
  year = 2024,
  institution = {{National Institute of Standards and Technology}},
  langid = {english}
}

@misc{schwartzRealityCheckNew2025,
  title = {Reality {{Check}}: {{A New Evaluation Ecosystem Is Necessary}} to {{Understand AI}}'s {{Real World Effects}}},
  shorttitle = {Reality {{Check}}},
  author = {Schwartz, Reva and Chowdhury, Rumman and Kundu, Akash and Frase, Heather and Fadaee, Marzieh and David, Tom and Waters, Gabriella and Taik, Afaf and Briggs, Morgan and Hall, Patrick and Jain, Shomik and Yee, Kyra and Thomas, Spencer and Bhandari, Sundeep and Duncan, Paul and Thompson, Andrew and Carlyle, Maya and Lu, Qinghua and Holmes, Matthew and Skeadas, Theodora},
  year = 2025,
  month = may,
  number = {arXiv:2505.18893},
  eprint = {2505.18893},
  primaryclass = {cs.CY},
  publisher = {arXiv},
  doi = {10.48550/arXiv.2505.18893},
  urldate = {2026-09-29},
  archiveprefix = {arXiv}
}

@article{schwartzWildThickWicked2026,
  title = {Wild, {{Thick}}, and {{Wicked}}: {{Situated Evidence}} on {{AI-In-Use}} for {{Decisions About Deploying AI Systems}}},
  shorttitle = {Wild, {{Thick}}, and {{Wicked}}},
  author = {Schwartz, Reva and Waters, Gabriella},
  year = 2026,
  month = sep,
  journal = {Social Science Computer Review},
  pages = {08944393261481511},
  publisher = {SAGE Publications Inc},
  issn = {0894-4393},
  doi = {10.1177/08944393261481511},
  urldate = {2026-09-29},
  langid = {english}
}

@misc{serranoLanguageModelsGuide2023,
  title = {Language {{Models}}: {{A Guide}} for the {{Perplexed}}},
  shorttitle = {Language {{Models}}},
  author = {Serrano, Sofia and Brumbaugh, Zander and Smith, Noah A.},
  year = 2023,
  month = nov,
  number = {arXiv:2311.17301},
  eprint = {2311.17301},
  primaryclass = {cs.CL},
  publisher = {arXiv},
  doi = {10.48550/arXiv.2311.17301},
  urldate = {2026-09-29},
  archiveprefix = {arXiv}
}

@inproceedings{sharmaUnderstandingSycophancyLanguage2024,
  title = {Towards Understanding Sycophancy in Language Models},
  booktitle = {The {{Twelfth International Conference}} on {{Learning Representations}}},
  author = {Sharma, Mrinank and Tong, Meg and Korbak, Tomasz and Duvenaud, David and Askell, Amanda and Bowman, Samuel R. and Durmus, Esin and {Hatfield-Dodds}, Zac and Johnston, Scott R. and Kravec, Shauna M. and Maxwell, Timothy and McCandlish, Sam and Ndousse, Kamal and Rausch, Oliver and Schiefer, Nicholas and Yan, Da and Zhang, Miranda and Perez, Ethan},
  year = 2024,
  note = {\url{https://openreview.net/forum?id=tvhaxkMKAn}}
}

@misc{smartTheoryValueAI2026,
  title = {Toward a {{Theory}} of {{Value}} in {{AI Alignment}}},
  author = {Smart, Andrew and Ahmed, Shazeda and Kay, Jackie and Tobin, Jimmy and Shrishak, Kris and Birhane, Abeba},
  year = 2026,
  month = aug,
  number = {arXiv:2608.10327},
  eprint = {2608.10327},
  primaryclass = {cs.AI},
  publisher = {arXiv},
  doi = {10.48550/arXiv.2608.10327},
  urldate = {2026-09-29},
  archiveprefix = {arXiv}
}

@inproceedings{stanczakSocietalFrameworksCan2026,
  title = {Societal {{Frameworks Can Improve LLM Alignment}}},
  booktitle = {Proceedings of the 2026 {{ACM Conference}} on {{Fairness}}, {{Accountability}}, and {{Transparency}}},
  author = {Sta{\'n}czak, Karolina and Meade, Nicholas and Bhatia, Mehar and Zhou, Hattie and B{\"o}ttinger, Konstantin and Barnes, Jeremy and Stanley, Jason and Papernot, Nicolas and Chapados, Nicolas and Therien, Denis and Lillicrap, Timothy P. and Marasovi{\'c}, Ana and Delacroix, Sylvie and Hadfield, Gillian K. and Reddy, Siva},
  year = 2026,
  month = jun,
  pages = {4491--4510},
  publisher = {Association for Computing Machinery},
  address = {New York, NY, USA},
  urldate = {2026-09-28},
  isbn = {979-8-4007-2596-8},
  doi = {10.1145/3805689.3806486}
}

@book{walzerThickThinMoral1994,
  title = {Thick and Thin: {{Moral}} Argument at Home and Abroad},
  author = {Walzer, Michael},
  year = 1994,
  publisher = {University of Notre Dame Press},
  address = {Notre Dame, IN}
}

@techreport{whitehouseofficeofscienceandtechnologypolicyBlueprintAIBill2022,
  title = {Blueprint for an {{AI}} Bill of Rights: {{Making}} Automated Systems Work for the {{American}} People},
  author = {{White House Office of Science and Technology Policy}},
  year = 2022,
  address = {Washington, DC},
  institution = {Executive Office of the President},
  note = {\url{https://bidenwhitehouse.archives.gov/ostp/ai-bill-of-rights/}}
}

@article{wienerMoralTechnicalConsequences1960,
  title = {Some {{Moral}} and {{Technical Consequences}} of {{Automation}}},
  author = {Wiener, Norbert},
  year = 1960,
  month = may,
  journal = {Science},
  volume = {131},
  number = {3410},
  pages = {1355--1358},
  publisher = {American Association for the Advancement of Science},
  doi = {10.1126/science.131.3410.1355},
  urldate = {2026-10-07}
}

@book{williamsEthicsLimitsPhilosophy1985,
  title = {Ethics and the {{Limits}} of {{Philosophy}}},
  author = {Williams, Bernard},
  year = 1985,
  publisher = {Harvard University Press},
  address = {Cambridge, MA},
  isbn = {978-0-674-26858-6},
  langid = {english}
}

@inproceedings{wilsonGenderRaceIntersectional2024,
  title = {Gender, {{Race}}, and {{Intersectional Bias}} in {{Resume Screening}} via {{Language Model Retrieval}}},
  booktitle = {Proceedings of the {{AAAI}}/{{ACM Conference}} on {{AI}}, {{Ethics}}, and {{Society}}},
  author = {Wilson, Kyra and Caliskan, Aylin},
  year = 2024,
  month = oct,
  volume = {7},
  pages = {1578--1590},
  issn = {3065-8365},
  doi = {10.1609/aies.v7i1.31748},
  urldate = {2026-11-02},
  publisher = {AAAI Press},
  copyright = {Copyright (c) 2024 Association for the Advancement of Artificial Intelligence},
  langid = {english}
}

@article{winnerArtifactsHavePolitics1980,
  title = {Do {{Artifacts Have Politics}}?},
  author = {Winner, Langdon},
  year = 1980,
  journal = {Daedalus},
  volume = {109},
  number = {1},
  pages = {121--136},
  publisher = {The MIT Press},
  issn = {0011-5266},
  urldate = {2026-10-07},
  note = {\url{https://www.jstor.org/stable/20024652}}
}

\end{document}